\documentstyle[prd,aps,preprint,tighten,epsf]{revtex}

\newcommand{\be}{\begin{equation}}
\newcommand{\ee}{\end{equation}}

\newcommand{\bd}{\begin{displaymath}}
\newcommand{\ed}{\end{displaymath}}

\newcommand{\bea}{\begin{eqnarray}}
\newcommand{\eea}{\end{eqnarray}}

\newcommand{\ba}[1]{\begin{array}{#1}}
\newcommand{\ea}{\end{array}}

\def\lvec#1{\setbox0=\hbox{$#1$}
    \setbox1=\hbox{$\scriptstyle\leftarrow$}
    #1\kern-\wd0\smash{
    \raise\ht0\hbox{$\raise1pt\hbox{$\scriptstyle\leftarrow$}$}}
    \kern-\wd1\kern\wd0}
\def\rvec#1{\setbox0=\hbox{$#1$}
    \setbox1=\hbox{$\scriptstyle\rightarrow$}
    #1\kern-\wd0\smash{
    \raise\ht0\hbox{$\raise1pt\hbox{$\scriptstyle\rightarrow$}$}}
    \kern-\wd1\kern\wd0}

\def\cancel#1#2{\ooalign{$\hfil#1\mkern1mu/\hfil$\crcr$#1#2$}}
\def\slash#1{\mathpalette\cancel{#1}}

\newcommand{\beq}{\begin{equation}}
\newcommand{\eeq}{\end{equation}}

\newcommand{\pslash}[1]{\rlap{/}\kern-0.8pt #1}
\newcommand{\lslash}{\rlap{/}\kern-0.0pt l}
\newcommand{\Dslash}{\rlap{/}\kern-2.0pt D}

\def\simge{
    \mathrel{\rlap{\raise 0.511ex
        \hbox{$>$}}{\lower 0.511ex \hbox{$\sim$}}}}
\def\simle{
    \mathrel{\rlap{\raise 0.511ex
        \hbox{$<$}}{\lower 0.511ex \hbox{$\sim$}}}}
\def\qw{Q^{\rm (w)}}
\def\qmp{Q^{\rm (mp)}}
\def\qbq{\langle \overline{q} q \rangle}
\def\mres{m_{\rm res}}
\def\dmqbq{\delta m_{\langle \, \overline{q} \, q \, \rangle}}

\begin{document}
\bibliographystyle{apsrev}



\preprint{CU-TP-1014, BNL-HET-01/10, RBRC-185}

\title{Chirality Correlation within Dirac Eigenvectors from Domain Wall Fermions} 

\author{
T.~Blum$^a$,
N.~Christ$^b$,
C.~Cristian$^b$,
C.~Dawson$^c$,
X.~Liao$^b$,
G.~Liu$^b$,
R.~Mawhinney$^b$,
L.~Wu$^b$,
Y.~Zhestkov$^b$}

\address{
\vspace{0.5in}
$^a$RIKEN-BNL Research Center,
Brookhaven National Laboratory,
Upton, NY 11973}

\address{$^b$Physics Department,
Columbia University,
New York, NY 10027}

\address{$^c$Physics Department,
Brookhaven National Laboratory,
Upton, NY 11973}

\date{May 6, 2001}
\maketitle

\begin{abstract}
In the dilute instanton gas model of the QCD vacuum, one expects a strong
spatial correlation between chirality and the maxima of the Dirac eigenvectors
with small eigenvalues.  Following Horvath, {\it et al.} we examine this
question using lattice gauge theory within the quenched approximation.
We extend the work of those authors by using weaker coupling, $\beta=6.0$,
larger lattices, $16^4$, and an improved fermion formulation, domain wall
fermions.  In contrast with this earlier work, we find a striking
correlation between the magnitude of the chirality density, 
$|\psi^\dagger(x)\gamma^5\psi(x)|$, and the normal density,
$\psi^\dagger(x)\psi(x)$, for the low-lying Dirac eigenvectors.
\end{abstract}


\newpage



\section{Introduction}
\label{sec:intro}

In a recent paper, Horvath, Isgur, McCune and Thacker~\cite{Horvath:2001ir}
suggest that an important test of various theoretical models of the 
QCD vacuum can be made by examining the degree to which the space-time
localization
of a low-lying eigenmode of the Dirac operator is correlated with 
non-zero chirality.  Such correlations are expected in the dilute
instanton gas model of the QCD vacuum.  

In this model, gauge configurations composed of widely separated 
instantons and anti-instantons support low-lying Dirac eigenvectors 
which are near superpositions of the zero modes that these 
(anti-)instantons would posses if in isolation.  These near zero modes 
are proposed \cite{Schafer:1998wv} to provide the non-zero density of 
Dirac eigenvalues, $\rho(\lambda)$ for vanishing eigenvalue $\lambda$, 
required by the Banks-Casher formula~\cite{Banks:1980yr} and the 
non-zero QCD chiral condensate, 
$\pi \rho(0) = -12\langle\overline{q} q\rangle = (245{\rm MeV})^3$,
where we give the value for $\langle\overline{q} q\rangle$ obtained
in a recent quenched calculation~\cite{Blum:2000kn} and the factor of
12 comes from our particular normalization conventions.   Since the 
isolated zero modes associated with an (anti-)instanton are entirely 
(left-)right-handed, one should expect a strong correlation of handedness 
with the locations at which $\psi_n^\dagger\psi_n(x)$ is large.

Thus, this class of instanton-based models of the QCD vacuum can be
tested by searching for such strong correlations between the 
chiral density $\psi_n^\dagger\gamma^5 \psi_n(x)$ and normal density 
$\psi_n^\dagger\psi_n(x)$ for the low-lying eigenmodes of the Dirac
operator on a configuration-by-configuration basis within a lattice QCD
calculation.  As emphasized by Horvath, {\it et al.} earlier arguments 
of Witten \cite{Witten:1979bc,Witten:1979vv} suggest that the large 
$N_c$ behavior of the $\eta^\prime$ mass may be inconsistent with the 
predictions of such an instanton model, providing additional motivation 
for such qualitative tests of the instanton picture.

In the above cited work of Horvath, {\it et al.}, the authors search for 
such correlations using 30 gauge configurations obtained in the quenched 
approximation on a $12^3 \times 24$ lattice volume with Wilson fermions 
and the Wilson gauge action with $\beta=5.7$.  In this paper, we extend
their study to smaller lattice spacing, 
$a^{-1}=1.922$ GeV, using a $16^4$ lattice volume which has a
similar physical size of $\approx 1.6$ Fermi.  Making use of configurations
that we had already analyzed, we discuss here results from two
different gauge actions.  The first is the standard Wilson action with 
$\beta=6.0$ for which we have 32 configurations and the second uses the Iwasaki
action with $\beta=2.6$, a value tuned to achieve the same lattice spacing.
The latter ensemble contains 55 configurations.

Finally we employ an improved fermionic action, using the domain wall 
formalism of Shamir \cite{Shamir:1993zy}.  Further references to this 
method, our implementation of this lattice fermion action and a complete
description of the notation used in this paper can be found in
Ref.~\cite{Blum:2000kn}. 

As is detailed below, using this finer lattice
spacing and improved fermion formulation, we find a very different
result from that seen in the earlier work of Horvath, {\it et al.}
The chirality of the $n^{th}$ eigenmode, $|{\psi}_n^\dagger(x)\gamma^5\psi_n(x)|$
is surprisingly close to maximum at those lattice sites $x$ where the
eigenmode is large.  In Section~\ref{sec:properties} we briefly review the
properties of the spectrum of the continuum Dirac operator and the closely 
related hermitian Dirac operator that we study. Section~\ref{sec:diagonal} 
explains the diagonalization procedure.  In Section~\ref{sec:results} we 
present our results and analysis of the domain wall fermion eigenvectors.  
Finally, Section~\ref{sec:conclusion} contains a brief conclusion.

\section{Properties of the Continuum Dirac Operator}
\label{sec:properties}

In the continuum, the Euclidean Dirac operator is usually written as a 
non-hermitian sum of an anti-hermitian operator $\slash{D}=\gamma^\mu D^\mu$
where $D^\mu$ is the usual gauge covariant derivative and a real 
mass term.  The combined operator, $\slash{D} +m$, is easy to 
analyze being the sum of commuting hermitian and anti-hermitian pieces.  
However, the standard Wilson and domain wall lattice Dirac operators 
are more complex because the so-called Wilson mass term also contains 
derivatives causing the hermitian and anti-hermitian parts of these lattice 
Dirac operators to fail to commute.  For the Wilson operator, $D_W$, this 
is easily remedied by working with the product $\gamma^5 D_W$ which is 
readily seen to be hermitian.  A similar construction is possible for
domain wall fermions~\cite{Furman:1995ky} where the product $\gamma_5 R_5 D$ 
is hermitian.  Here $D$ is the domain wall Dirac operator, given for 
example in Eq.~1 of Ref.~\cite{Blum:2000kn}, and $R_5$ the reflection in 
the midplane in the fifth dimension, $s \rightarrow L_s-1-s$.

Our treatment differs from that of Horvath, {\it et al.} in that we work
with the hermitian operator, finding real eigenvalues and their corresponding
eigenvectors while the authors of Ref.~\cite{Horvath:2001ir} determine the
complex eigenvalues of the non-hermitian Wilson operator and their
corresponding eigenvectors.  In order to compare our two results, we 
now review the relation between the eigenvectors of corresponding continuum 
operators: $\slash{D} +m$ and $\gamma^5(\slash{D} +m)$.

We begin by recalling the basic properties of the eigenmodes of the 
Dirac operator $\slash{D}$ and the hermitian operator 
$\gamma^5 \slash{D}$ in the continuum and then discuss properties of 
the local chiral density which is the focus of this paper.
Since $\left\{ \gamma^5 , \slash{D} \right\} = 0$ and 
$\slash{D}^{\dagger} =  - \slash{D} \,$,
the spectrum of $\slash{D} $ consists of pairs of
eigenvectors $\psi_{\lambda}(x)$ and 
$\psi_{-\lambda}(x)=\gamma^5 \psi_\lambda(x)$ with imaginary eigenvalues 
$\pm i \lambda$.  For $\lambda=0$ the zero modes, 
$\{\psi_{0,i}\}_{1 \le i \le N}$, can also be chosen 
eigenstates of $\gamma^5$, $\gamma^5 \psi_{0,i} = \pm \psi_{0,i}$ since
$\left[ \gamma^5 , \slash{D} \right] \psi_{0,i} = 0$.  Here we have added
the extra subscript $i$ to the zero modes to allow for the possibility that
more than one eigenvector with zero eigenvalue exists.  The Atiyah-Singer 
index theorem requires that the number of zero modes, $N$ satisfy,
$N\ge |\nu|$ where $\nu$ is the integer winding number of the gauge field
($\nu=\dots -2,-1,0,1,2,\dots$).

Clearly, zero modes of $\slash{D}$ are also zero modes of 
$\gamma^5 \slash{D}$.  The remaining eigenvectors of $\gamma^5(\slash{D}+m)$ 
can be easily related to those of $\slash{D}+m$ by diagonalizing 
the former on the two-dimensional subspace spanned by the basis  
$\left\{ \psi_{\lambda}, \psi_{-\lambda} \right\}$:
\bea
\gamma^5 \left( \slash{D} + m \right)
\left( \begin{array}{c}
a\\
b\\
\end{array} \right)
&=&
\left( \begin{array}{cc}
0 & m - i\lambda \\
m+i\lambda & 0   \\
\end{array} \right)
\left( \begin{array}{c}
a\\
b\\
\end{array} \right)
\\ \nonumber
&=&
\lambda_H
\left(\begin{array}{c}
a\\
b\\
\end{array}\right).
\eea
The eigenvalues are easily seen to be
\be
\lambda_H = \pm \left[ \lambda^2 + m^2 \right]^{\frac{1}{2}} \, .
\ee
The normalized eigenvectors, $\psi_{H,\lambda_H}$ are
\be
\psi_{H,\pm|\lambda_H|}  =
\frac{1}{\sqrt{2}\left( \lambda^2  + m^2 \right)^{\frac{1}{4}} }
    \left[ \left(m-i\lambda \right)^{\frac{1}{2}}\psi_\lambda
       \pm \left(m+i\lambda \right)^{\frac{1}{2}}\psi_{-\lambda} \right] \,.
\label{eq:eig}
\ee

Next we compare the chiral density as determined by these two
sets of eigenvectors.  For zero modes:
\be
\psi_{0,i}^\dagger(x)\gamma^5\psi_{0,i}(x)
   = \pm \psi_{0,i}^\dagger(x)\psi_{0,i}(x),
\ee
the largest value possible, for both the hermitian and non-hermitian
operators.  For the other, paired eigenvectors and the non-hermitian 
operator, the identity $\psi_{-\lambda}(x)=\gamma^5\psi_\lambda(x)$ 
implies that product
\be 
\psi_\lambda^\dagger(x)\gamma^5\psi_\lambda(x)
=\psi_{-\lambda}^\dagger(x)\gamma^5\psi_{-\lambda}(x)
= \chi(x)
\label{eq:chi}
\ee
is the same for each eigenvector $\psi_{\pm\lambda}$ in the pair.
Similarly, the norm of the wave function at each point in space-time is
the same for each pair of eigenvectors
\be 
\psi_\lambda^\dagger(x)\psi_\lambda(x)
=\psi_{-\lambda}^\dagger(x)\psi_{-\lambda}(x)
= \omega(x)
\label{eq:omega}
\ee

This is not true for the paired eigenvectors of the hermitian
operator, where using Eq.~\ref{eq:eig} we find:
\bea
\psi_{H,\lambda_H}^\dagger(x) \psi_{H,\lambda_H}(x)
  &=& w(x) \pm \epsilon \chi(x) = \omega_H(x),
\\ \nonumber
\psi_{H,\lambda_H}(x)\gamma^5 \psi_{H,\lambda_H}(x)
&=&
\chi(x) \pm \epsilon \omega(x) = \chi_H(x),
\label{eq:chi_hermitian}
\eea
where
\bea
\epsilon &=& \frac{m}{\sqrt{\lambda^2 + m^2}}
\label{eq:epsilon}
\eea
and $\lambda = \pm |\lambda|$.
In the chiral limit $\epsilon$ vanishes, and the local chiral densities 
of the anti-hermitian and hermitian Dirac operators coincide.  

The above analysis should describe the properties of the low-lying
eigenstates of the non-hermitian Wilson operator in the continuum limit.
Thus, we conclude that the difference between the evaluation
of the chiral density for the eigenvectors of the non-hermitian Wilson
Dirac operator and those of the hermitian domain wall Dirac operator
should agree in the continuum limit for those eigenvectors for which the
parameter $\epsilon$ is small.  Thus, the results presented in this 
paper and those of Horvath, {\it et al.} address the same continuum question
and can be compared.  Since in our calculation the explicit fermion mass is 
very small ($m_f=0$ for the Wilson action and $m_f=5 \cdot 10^{-4}$ for the Iwasaki
action) we expect that for most modes the parameter $\epsilon$ will also
be very small making the normal and chiral densities the same for both 
operators.  The only exception to this conclusion arises for the 
very smallest eigenvalues, $\lambda \approx 10^{-3}$ where $\epsilon$ may 
be of order one due to the residual mass coming from mixing between 
the walls, $m_{\rm res} \approx 10^{-3}$ for the Wilson 
action~\cite{Blum:2000kn,AliKhan:2000iv} and $m_{\rm res} \approx 10^{-4}$ 
for the Iwasaki action~\cite{AliKhan:2000iv,us:2001}.)

\section{Diagonalization Method}
\label{sec:diagonal}

Using the conjugate-gradient method proposed by
Kalkreuter and Simma~\cite{Kalkreuter:1996mm}, we calculate the 19 lowest 
eigenvalues and eigenvectors of $D_H$.   This corresponds to the range 
$0\simle \lambda \simle 200$(MeV).   In this section we describe
our implementation of their method\footnote{Code provided
by Robert Edwards forms the basis for the computer program used here.}.

Following Kalkreuter and Simma, we use the conjugate gradient method to 
minimize the Ritz functional,
\be
\mu(\Psi)= {\langle\Psi|(D_H)^2|\Psi\rangle \over \langle\Psi|\Psi\rangle}
\ee
on a sequence of subspaces orthogonal to the eigenvectors corresponding 
to the previously identified minima.  When we have obtained our intended 
$N_{\rm max} = 19$ eigenvectors, we then evaluate the 
$N_{\rm max} \times N_{\rm max}$ matrix $\langle\Psi_k|(D_H)^2|\Psi_j\rangle$.  
This small matrix is then diagonalized by a Jacobi method and the 
resulting transformation used to improve the initial set of vectors 
$\{\Psi_k\}_{0\le k < N_{\rm max}}$.  This combination of 
conjugate gradient minimizations and small matrix diagonalizations is 
repeated until a sufficiently accurate result has been obtained. 

This method requires the choice of two stopping criteria.  For the 
sequence of $N_{\rm max}$ conjugate gradient minimizations of 
$\mu(\Psi_k)$ we iterate each conjugate gradient procedure until the 
norm squared of the gradient of $\mu(\Psi_k)$ with respect to 
$\Psi_k$ has decreased from its initial value by a factor of 10.  
This choice of $\gamma=0.1$, in the notation of Kalkreuter and Simma, 
was recommended by those authors and, after some testing, appeared 
to be a good choice for us.  As a criterion to stop the outer 
iteration over $N_{\rm max}$ conjugate gradient minimizations 
followed by small matrix diagonalization, we required that the 
relative change in each of the eigenvalues be less than $10^{-7}$ 
after the final step in the method.

Since we are actually interested in the small eigenvalues and 
corresponding eigenvectors of $D_H$, we performed a follow-on step 
after the above procedure has yielded a best set of eigenvectors 
and eigenvalues of $D_H^2$.  In this step, we used the Jacobi method
again to diagonalize the $19 \times 19$ matrix, 
$\langle\Psi_k|D_H|\Psi_j\rangle$ and the resulting transformation 
to rotate the original basis of 19 eigenvectors into what should be 
eigenvectors of $D_H$.

However, if the $19^{th}$ and $20^{th}$ eigenvalues of $D_{H}^2$ are 
nearly degenerate but the corresponding eigenvalues of $D_{H}$ have
opposite signs, then this 19-dimensional subspace will in general 
not be invariant under the application of $D_{H}$.  The $19^{th}$
eigenvector of $D_H^2$ can be an arbitrary mixture of the $19^{th}$ and
$20^{th}$ eigenvectors of $D_H$.  If this happens, we expect to see one 
``spurious'' and 18 valid eigenvectors of $D_{H}$ after the Jacobi step. 
The spurious eigenvector will be orthogonal to the other 18 and the
magnitude of the corresponding eigenvalue will be distributed 
arbitrarily between a correct $19^{th}$ value and zero.

One way to identify such a spurious eigenvalue is to look for an eigenvalue
of $D_{H}$ whose square does not agree with any of the 19 eigenvalues of 
$D_{H}^2$.  In practice we order the eigenvalues found for both $D_H$
and $D_H^2$ and consider 19 possible matches obtained by discarding one
of the eigenvalues of $D_H$ and comparing the squares of the remaining
18 with the lowest 18 eigenvectors of $D_H^2$.  The case with the largest
sum of squares of relative differences identifies the spurious eigenvalue.
This algorithm produces 18 ``good'' eigenvalues and discards the largest
eigenvalue in the case that no spurious eigenvector is present.  We use
this method to determine the 18 eigenvectors analyzed in this paper.

A second method that in addition involves the eigenvector information 
evaluates the anti-commutator identity satisfied by $D_{H}$,   
\begin{equation}
  \{ \Gamma_5, D_H \} =  2m_f \qw + 2 \qmp,
\label{eq:sym_constraint}
\end{equation}
where $\qw$ and $\qmp$ are pseudoscalar densities defined on the pairs of
planes at $s=(0,L_s-1)$ and $s=(L_s/2-1,L_s/2)$, respectively, with 
precise definitions given in Ref.~\cite{Blum:2000kn}.  We compute the 
diagonal elements of this matrix equation in our basis of eigenvectors 
of $D_H$:
\be
\Lambda_n\langle\Psi_{n}|\Gamma_5|\Psi_{n}\rangle
         = \langle\Psi_{n}|\Bigl[m_f\qw + \qmp \Bigr]|\Psi_{n}\rangle.
\label{eq:ks_test}
\ee
If for one or more eigenvectors the fractional difference of the left- 
and right-hand sides of Eq.~\ref{eq:ks_test} is larger than 0.12 we
discard the eigenvector producing the largest difference.  If all
eigenvectors give differences lying below this value, we discard the
largest eigenvalue.  The fraction 0.12 is a somewhat arbitrary dividing 
point that separates the few large discrepancies from the majority with 
a much smaller difference.

Comparison of these two criteria shows disagreement on 4 of the
32 Wilson and 6 of the 55 Iwasaki configurations.  However, in each
of those cases one method discards the nineteenth eigenvalue while
the other identifies either the seventeenth or eighteenth as spurious.
Since the eigenvectors of these largest eigenvalues are likely the
least well determined, we view this as satisfactory and expect that
the difficulties from this imprecise identification of ``spurious'' 
eigenvectors enter our sample of eigenvectors on the level of a few 
percent or less.

\section{Analysis and Results}
\label{sec:results}

In this section we examine some of the properties of the 18 lowest eigenvalues, 
$\Lambda_{H,i}$ and the corresponding eigenvectors $|\Lambda_{H,i}\rangle$
determined by the procedure described above, 
\be
D_H |\Lambda_{H,i}\rangle =\Lambda_{H,i}|\Lambda_{H,i}\rangle \quad (0 \le i < 18).
\ee
The connection between these eigenvectors and eigenvalues and those of the 
usual 4-dimensional Dirac operator was discussed in detail 
in~\cite{Blum:2000kn}.  The 5-dimensional wave functions, $\Psi_{H,i}(x,s)$ 
are expected to represent the wave functions of 4-dimensional 
Dirac eigenfunctions with an added exponential dependence on the fifth
coordinate $s$, causing $|\Psi_{H,i}(x,s)|^2$ to fall rapidly as $s$ moves
away from the 4-dimensional planes $s=0$ and $s=L_s-1$, as shown, for example
in Figs.~29 and 30 of Ref~\cite{Blum:2000kn}.  Likewise the eigenvalues 
$\Lambda_{H,i}$ should correspond to 4-dimensional Dirac eigenvalues.

For our conventions the components bound to the $s=0$ wall are predominately
left-handed while those concentrated on the $s=L_s-1$ wall are right-handed.
It is important to recall that for a lattice with finite $L_s$, a 
left-handed mode can propagate with a very small probability to the 
right-handed wall, and visa-versa, giving a small breaking of chiral 
symmetry which, for low energy phenomena, can be described by a residual 
quark mass, $m_{\rm res}$.

Our analysis is based primarily on 55 quenched gauge configurations 
generated with the Iwasaki gauge action~\cite{Iwasaki:1983ck} at $\beta=2.6$ 
and 4-dimensional lattice volume $16^4$. We also examine 32 configurations 
generated with the standard Wilson action at $\beta=6.0$ and the same 
volume.  These two gauge actions have been choosen to correspond to
approximately the same lattice spacing: $a^{-1} \approx 2$ GeV.  
The gauge configurations that we analyze are separated by 
2000 sweeps of our updating algorithm, a simple two-subgroup heat-bath 
update of each link.  The number of sites in the fifth dimension for 
the domain wall Dirac operator is $L_s=16$, and the domain wall height 
is $M_5=1.8$.  For input quark mass we take $m_f=0$ for the Wilson
action and $m_f=5\cdot 10^{-4}$ for the Iwasaki case.  This small,
non-zero value of $m_f$ is used in the Iwasaki case to avoid the very
slow convergence of the Kalkreuter Simma method that we found when
using $m_f=0$ for that action.  Finally, the 
residual quark mass for these couplings is quite small, 
$m_{\rm res}\approx 0.0001$ for Iwasaki\cite{AliKhan:2000iv,us:2001}, 
and $m_{\rm res} \approx 0.001$ for Wilson\cite{AliKhan:2000iv,Blum:2000kn} 
which correspond to 0.4 and 4 MeV, respectively.

\begin{figure}
\epsfxsize=\hsize
\epsfbox{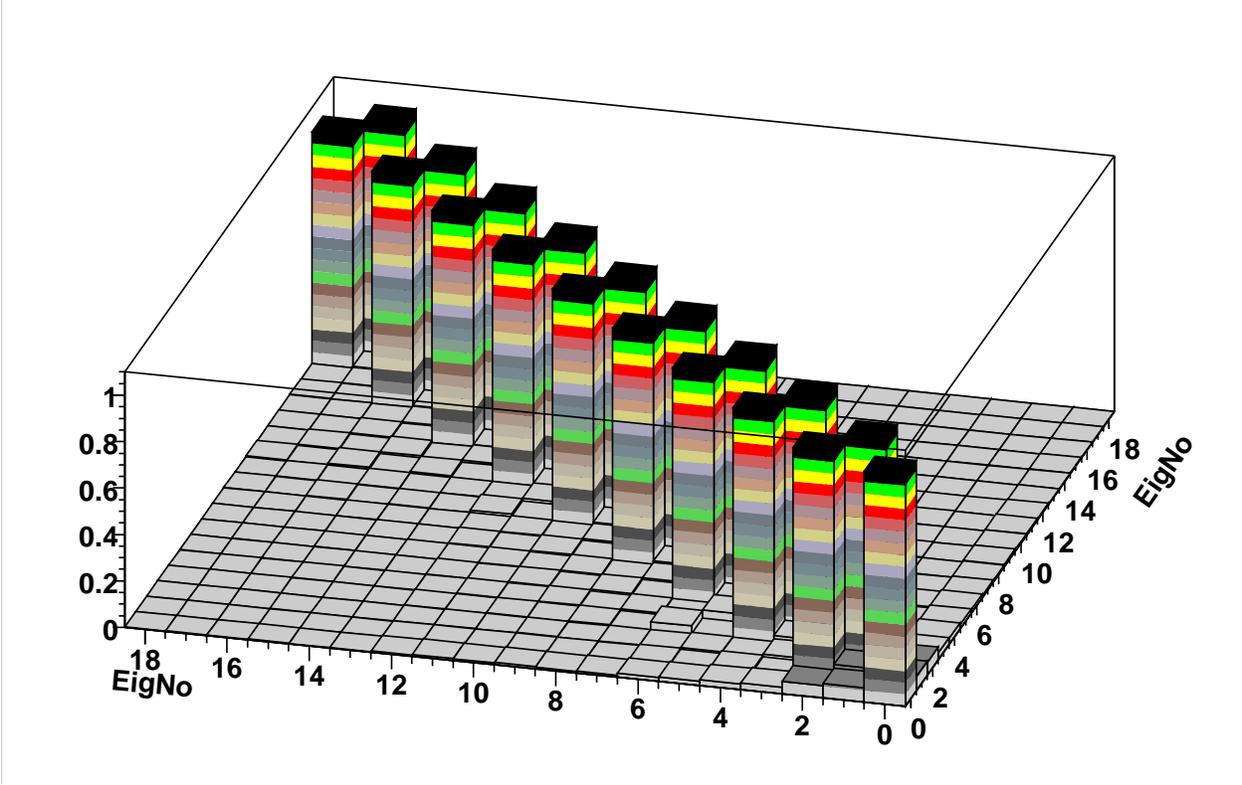}
\vskip .25in
\caption{The magnitude of the matrix elements 
$\langle \Lambda_{H,i}|\Gamma^5|\Lambda_{H,j}\rangle$ evaluated
on a ``simple'' configuration.  The pattern seen is close to that
expected for matrix elements of $\gamma^5$ in the continuum theory:
The single near zero mode is an an approximate eigenstate of $\Gamma^5$ 
while the non-zero modes come in $\pm \Lambda_H$ pairs related 
by $\Gamma^5 \Psi_{\Lambda_H} = -\Psi_{\Lambda_H}$.}
\label{fig:simple_lego}
\end{figure}

\begin{figure}
\epsfxsize=\hsize
\epsfbox{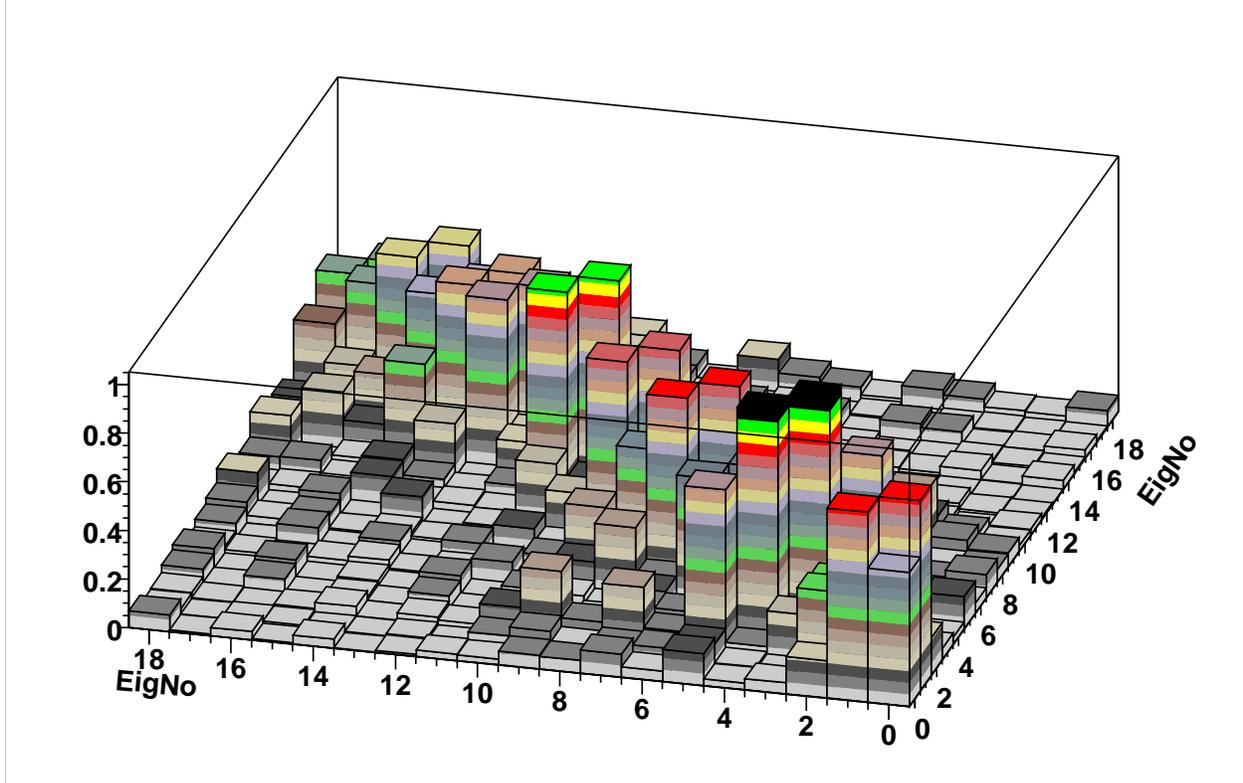}
\vskip .25in
\caption{The magnitude of the matrix elements 
$\langle \Lambda_{H,i}|\Gamma^5|\Lambda_{H,j}\rangle$ evaluated
on a ``complex'' configuration.  For the Wilson gauge action more than one-half
of the 32 configurations look like this while for the Iwasaki action
only one in ten has this complex structure with the remaining 90\%
appearing similar to Fig.~\protect\ref{fig:simple_lego}.}
\label{fig:complex_lego}
\end{figure}

To begin, let us examine the integrated or global chiral structure of 
$D_H$.  We plot the magnitude of 
$\langle \Lambda_{H,i}| \Gamma^5|\Lambda_{H,j}\rangle$ in 
Figs.~\ref{fig:simple_lego} and~\ref{fig:complex_lego} for a typical 
``simple'' and ``complex'' configuration.  The matrix
$\Gamma^5$ represents the physical $\gamma^5$ matrix in the domain
wall fermion formalism~\cite{Furman:1995ky}:
\be
\langle \Lambda_{H,i}| \Gamma^5|\Lambda_{H,j}\rangle =
      \sum_{x \in V} \sum_{s=0}^{L_s-1} 
            \Psi_{\Lambda_{H,i}}^\dagger(x,s){\rm sign}(s-(L_s-1)/2) 
                         \Psi_{\Lambda_{H,j}}(x,s)\,,
\ee
where $V$ is the 4-dimensional space-time volume.

The pattern seen in Fig.~\ref{fig:simple_lego} for the simple configuration
is precisely the chiral structure expected in the continuum.  The single,
diagonal element corresponding to $|\Lambda_{H,0}\rangle$ represents
a zero mode which is an eigenstate of $\gamma^5$ with eigenvalue $+1$.
(We will refer to such modes as ``near zero modes'' since for our choice
of parameters, their eigenvalues are not precisely zero.)
In our entire sample of Iwasaki and Wilson configurations, all such near zero
modes have very small eigenvalues, $\Lambda \approx 10^{-3} - 10^{-4}$,
and all are either right-handed or left-handed within a given gauge
configuration.  Note this behavior is a natural consequence of the
Atiyah-Singer theorem which requires an excess of right-handed to 
left-handed zero modes equal to the winding number 
of the background gauge configuration.  This determines a minimum number 
of zero modes, all with the same chirality.  The presence of additional 
zero modes would imply added constraints on the gauge background, corresponding
to a set of zero measure if our near zero modes had a precisely zero eigenvalue. 
The remaining eigenvectors are grouped into pairs connected by $\gamma^5$ precisely
as expected for the continuum Dirac operator in the limit of vanishing
mass.

These simple continuum expectations are not satisfied by the complex 
configurations, such as that shown in Fig.~\ref{fig:complex_lego}.   Of course,
such configurations must be present for finite $L_s$ and finite lattice
spacing.  As the gauge configurations change continuously from one
winding number to another, a plot of the sort shown in these figures
must also change continuously and hence cannot always have the simple
structure of Fig.~\ref{fig:simple_lego}.  While for the Wilson gauge 
action somewhat more than half of the configurations show the complex
pattern of Fig.~\ref{fig:complex_lego}, for the better-behaved Iwasaki case,
this fraction has dropped to 10\%.  It should be emphasized that all
the low-lying eigenvectors studied, both complex and simple ones, fall
off rapidly away from the walls with the minimum magnitude of the wave 
function between the walls falling at least a factor thirty below its value
on the two physical boundaries.

Such gauge configurations in which the winding number is changing can be 
associated with zero modes of the 4-dimensional Wilson Dirac operator 
with mass equal to $-M_5$~\cite{Narayanan:1995gw,Edwards:1998sh}.
Numerical simulations~\cite{Edwards:1999bm} suggest that the density 
of such 4-dimensional Dirac operator zero modes decreases exponentially
with the exponential of the coupling, $\rho \sim e^{-c/\sqrt{a}}$, so
that such effects may vanish rapidly as the continuum limit is 
approached.

Close to the continuum limit, the typical gauge configuration 
will be sufficiently continuous that its winding number 
can be identified~\cite{Luscher:1982zq}.  As the winding 
number changes one expects that localized, rapidly
changing gauge fields will appear and small dislocations, on the
scale of a very few lattice spacings will appear or disappear.
It is natural to speculate that such configurations produce the
complex $\Gamma^5$ matrix elements of Fig.~\ref{fig:complex_lego}
and the non-zero density of 4-dimensional Dirac zero modes 
described above.  The comparison of Iwasaki and Wilson results
suggests that while such configurations are quite common for 
the Wilson gauge action when $a^{-1} \approx 2$ GeV, they are
dramatically suppressed under similar circumstances by the form 
of the action proposed by Iwasaki.  This general topic is the 
subject of much current 
research~\cite{Berruto:2000fx,Edwards:2000qv,Hernandez:2000iw,Shamir:2000cf}. 

If we are to systematically evaluate the domain wall fermion, QCD path 
integral, we must include all configurations in our analysis.  Although 
we are explicitly examining the small eigenmodes of the Dirac operator, 
the eigenvalues and eigenfunctions represent a complex mixture of both
long-distance and short-distance physics.  While we are able to explicitly
focus on small Dirac eigenvalues, the background gauge fields contain
the full spectrum of short- and long-distance fluctuations.  This is 
immediately demonstrated by the potential importance of small, very 
short-distance dislocations in the gauge configuration on quantities 
we are examining.  However, we should also expect the more conventional 
short distance effects of wave function and mass renormalization, 
including short-distance contributions to the residual mass, to 
influence the low lying eigenvalues and eigenvectors analyzed here.
A brief discussion of these effects on the eigenvalue spectrum can
be found in Section VI C of Ref.~\cite{Blum:2000kn}.
We speculate that these complex configurations, which do not
behave as is expected for smooth, continuum gauge fields, represent
such short-distance effects.  However, we should look carefully to see 
if configurations of the complex type introduce further chiral symmetry 
breaking at low energy beyond the simple residual mass described above.  
To date we have not recognized such effects.

\begin{figure}
\epsfxsize=\hsize
\epsfbox{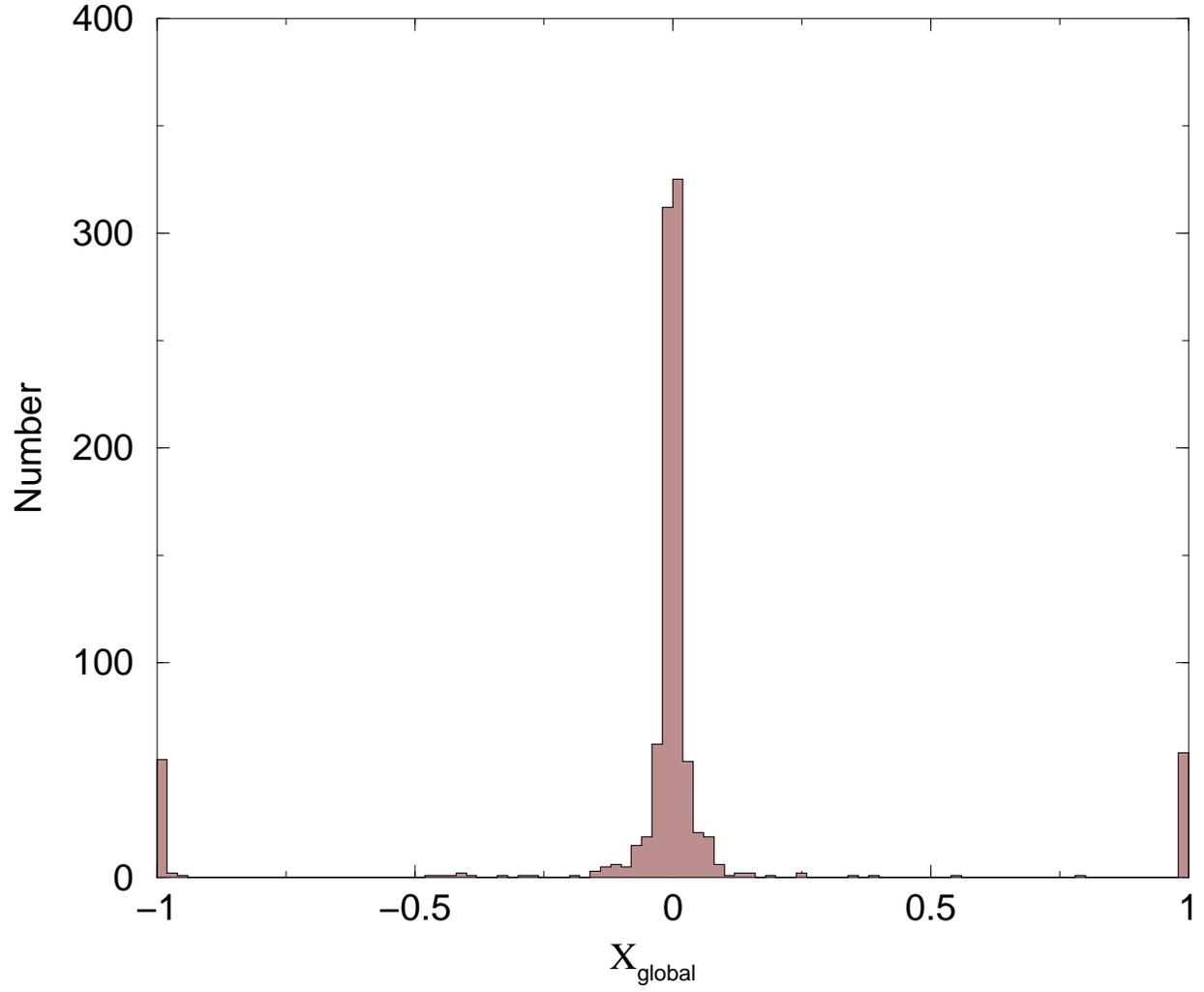}
\vskip .25in
\caption{The distribution of global chirality of the eigenvectors of 
$D_H$ evaluated at zero input quark mass, $m_f=0$.}
\label{fig:global_chirality}
\end{figure}

The global chirality of our eigenvectors is summarized in 
Fig.~\ref{fig:global_chirality}. The distribution shows a large
narrow peak around zero corresponding to the non-zero modes 
and two smaller, but also narrow, peaks at $\pm 1$ corresponding to 
near zero modes.  From these figures it is clear that we can easily 
distinguish near zero modes from non-zero modes, except for the handful 
of outliers with chirality neither close to zero nor $\pm 1$.

\begin{figure}
\epsfxsize=\hsize
\epsfbox{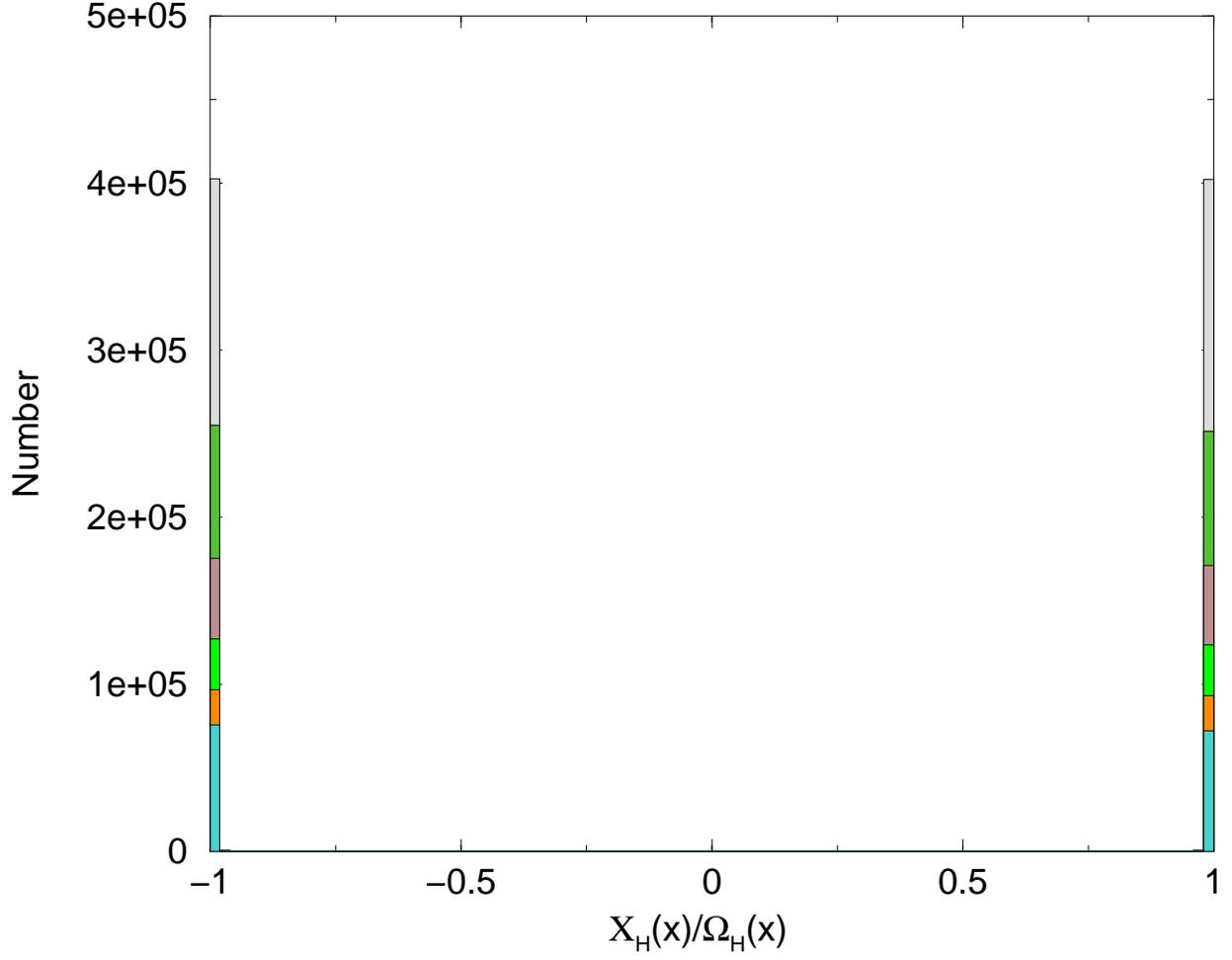}
\vskip .25in
\caption{The distribution of local chirality $X_H(x)/\Omega_H(x)$
of near zero mode eigenvectors of $D_H$ on the Iwasaki ensemble of gauge 
fields for sites where $\Psi^\dagger(x)\Psi(x)$ is greater than an 
arbitrary imposed cut.  The different cuts correspond to keeping between 1 
and 8\% of the total sites in the space-time lattice.}
\label{fig:local_chirality_zm}
\end{figure}

\begin{figure}
\epsfxsize=\hsize
\epsfbox{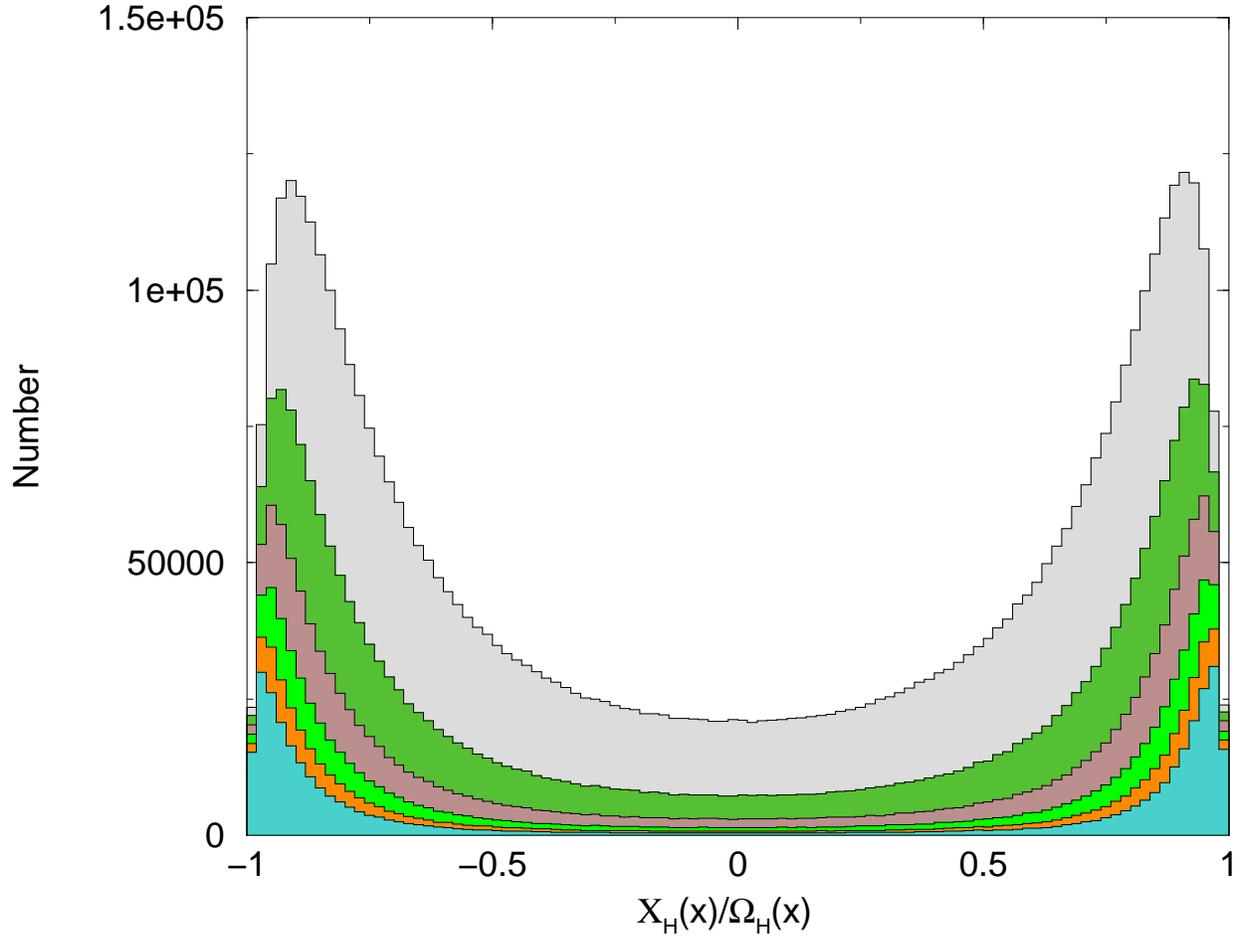}
\vskip .25in
\caption{The same quantities as in Fig.~\protect\ref{fig:local_chirality_zm}, 
but for non-zero mode eigenvectors and again for the Iwasaki action.  The 
double-peak structure is a feature expected in instanton-dominated models 
of the QCD vacuum.}
\label{fig:local_chirality_nzm}
\end{figure}

Next we examine the distribution of local chirality, $X_H(x)/\Omega_H(x)$, 
which is shown in Figs.~\ref{fig:local_chirality_zm} 
and~\ref{fig:local_chirality_nzm} for near zero modes and non-zero modes, 
respectively.  Here the quantities $X_H(x)$ and $\Omega_H(x)$ are the 
generalizations of the continuum quantities $\chi_H(x)$ and $\omega_H(x)$ 
defined in Eq.~\ref{eq:chi_hermitian}, to the case of domain wall fermions:
\bea
\Omega_H(x) &=& \sum_{0 \le s < L_s} \Psi_{\Lambda_{H,i}}(x,s)^\dagger
                      \Psi_{\Lambda_{H,i}}(x,s) \nonumber \\
X_H(x)      &=& \sum_{0 \le s < L_s} {\rm sign}(s-(L_s-1)/2)
                      \Psi_{\Lambda_{H,i}}(x,s)^\dagger
                        \Psi_{\Lambda_{H,i}}(x,s)
\eea
The six histograms superimposed in Figs.~\ref{fig:local_chirality_zm} 
and~\ref{fig:local_chirality_nzm} correspond to histograms of the quantity
$X_H(x)/\Omega_H(x)$ evaluated only at those sites where the normal density
$\Omega_H(x)$ lies above a specified cut, for each of the eighteen 
eigenvectors that we have determined.  The six cuts displayed correspond
to the conditions: $\Omega_H(x)\cdot 10^{5} > 3, 4, 5, 6, 7$ 
and 8 which include 8.6, 4.3, 2.4, 1.5 1.0 and 0.7\% of the sites in the space-time 
lattice, respectively.   These cuts correspond to sampling the local chirality, 
on average, from sites that account for 28, 19, 13, 10, 8, and 6\%, 
respectively, of the total probability density of each eigenvector.
In Table~\ref{tab:cuts} we give these numbers with more precision for
both the Iwasaki and Wilson cases.

The near zero mode distributions are sharply peaked at $\pm 1$ for all cuts, 
as expected.  The non-zero mode distributions are also clearly double-peaked, 
with peaks centered approximately around $\pm .95\to \pm.8$, depending on 
the size of the cut.  The distributions fill in between the peaks as more 
sites are sampled.  Thus, our data show that definite chirality is strongly 
correlated with local maxima of both near zero and non-zero eigenmodes.  The 
latter is certainly consistent with an instanton-dominated vacuum 
picture and is in disagreement with the recent work of Horvath, {\it et al.}.

\begin{figure}
\epsfxsize=\hsize
\epsfbox{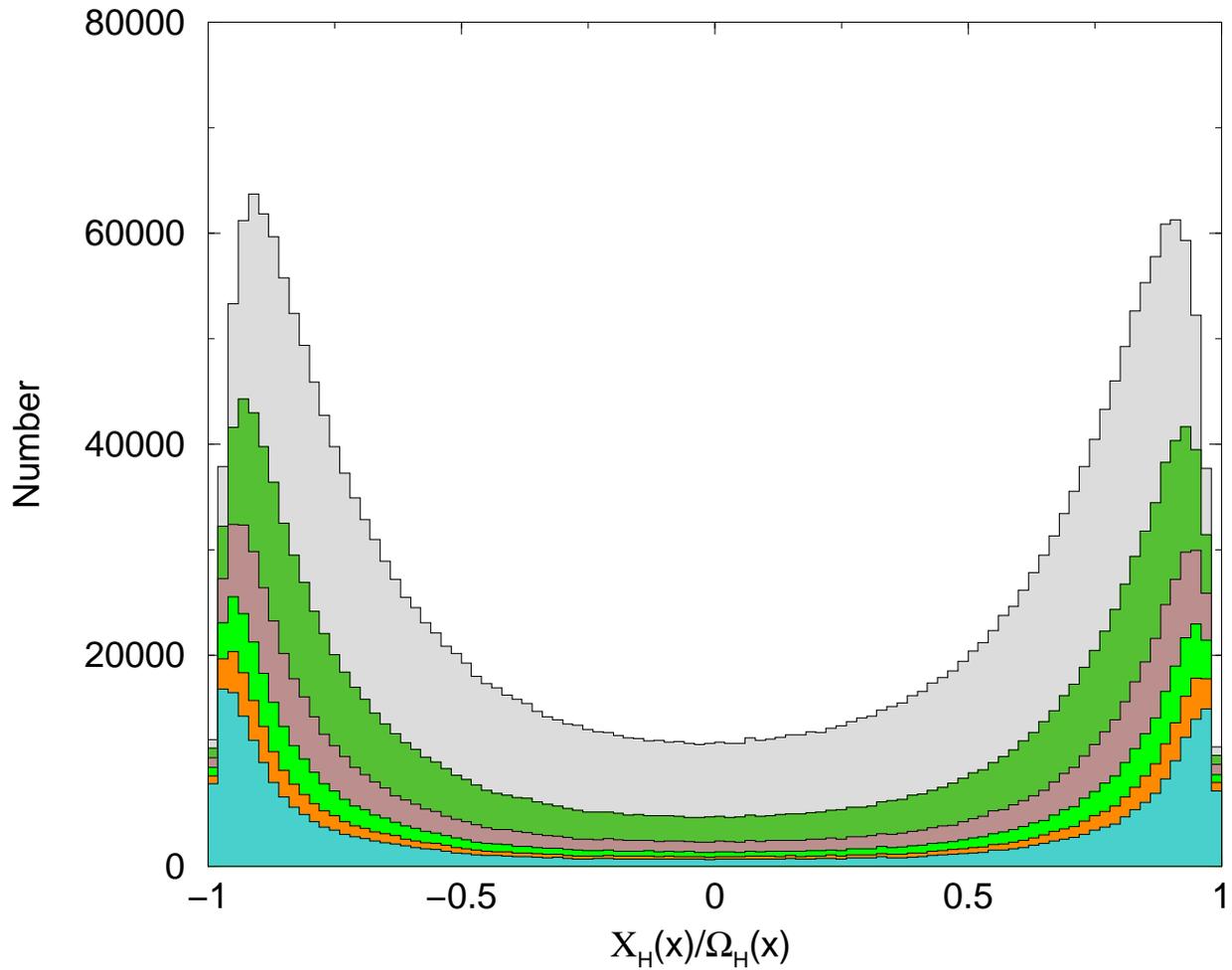}
\vskip .25in
\caption{The same quantities as in Fig.~\protect\ref{fig:local_chirality_nzm}, 
except now the Wilson gauge configurations are examined.  The double-peak 
structure is very similar to that found in the Iwasaki case.}
\label{fig:local_chirality_wilson}
\end{figure}

We also note that the strong correlation between chiral and normal density
for our non-zero mode eigenvectors is also evident if we use the definition 
of local chirality in~\cite{Horvath:2001ir} instead of the ratio
$X_H(x)/\Omega_H(x)$.  In Fig.~\ref{fig:local_chirality_wilson} we 
show a similar set of histograms for the Wilson gauge action.  Clearly the 
same strong correlation between chiral and normal density is seen for 
this case as well. 

\begin{figure}
\epsfxsize=\hsize
\epsfbox{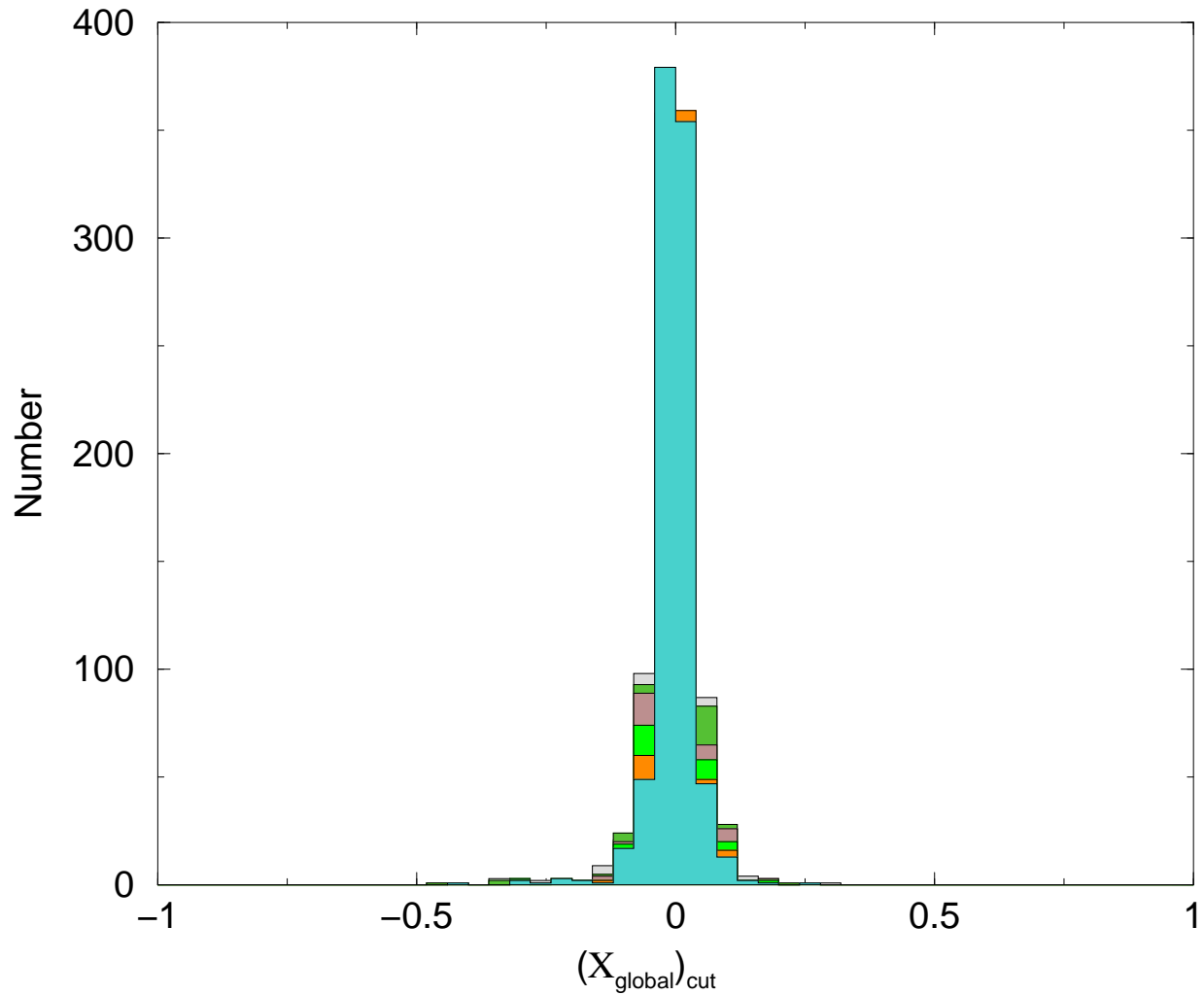}
\vskip .25in
\caption{The distribution of chiral density for non-zero mode eigenvectors of 
$D_H$ summed over those lattice sites obeying the cuts imposed on $\Omega_H(x)$ 
as described in the text.  As the cut is increased, the width of 
the distribution becomes more narrow.  The distributions indicate that 
on average, chirality for a single eigenvector is evenly split into 
left- and right-handed lumps.}
\label{fig:cut_global_chirality}
\end{figure}

We now discuss two important consistency checks on this potentially
interesting result.  First we determine the distribution of ``global'' 
chirality of each eigenvector when computed by including only those lattice 
sites selected by our cuts on $\Omega_H(x)$.   We expect this distribution 
to be similar to that found when summing over all sites in the lattice, 
with each non-zero mode showing approximately zero total chirality.
The results are shown in Fig.~\ref{fig:cut_global_chirality}.  While considerably
broader than the distribution seen for the global chirality, it is still
strongly peaked around zero with the sharpness of the peak decreasing
as the cut on $\Omega_H(x)$ is made more stringent.  This broadening is
easily understood as the statistical effect of examining the sum over
a smaller number of space-time points.  Over-all, 
Fig.~\ref{fig:cut_global_chirality} is quite consistent with the expectation
that chirality of the non-zero mode eigenvectors is quite evenly split between
left- and right-handed lumps.

\begin{figure}
\epsfxsize=\hsize
\epsfbox{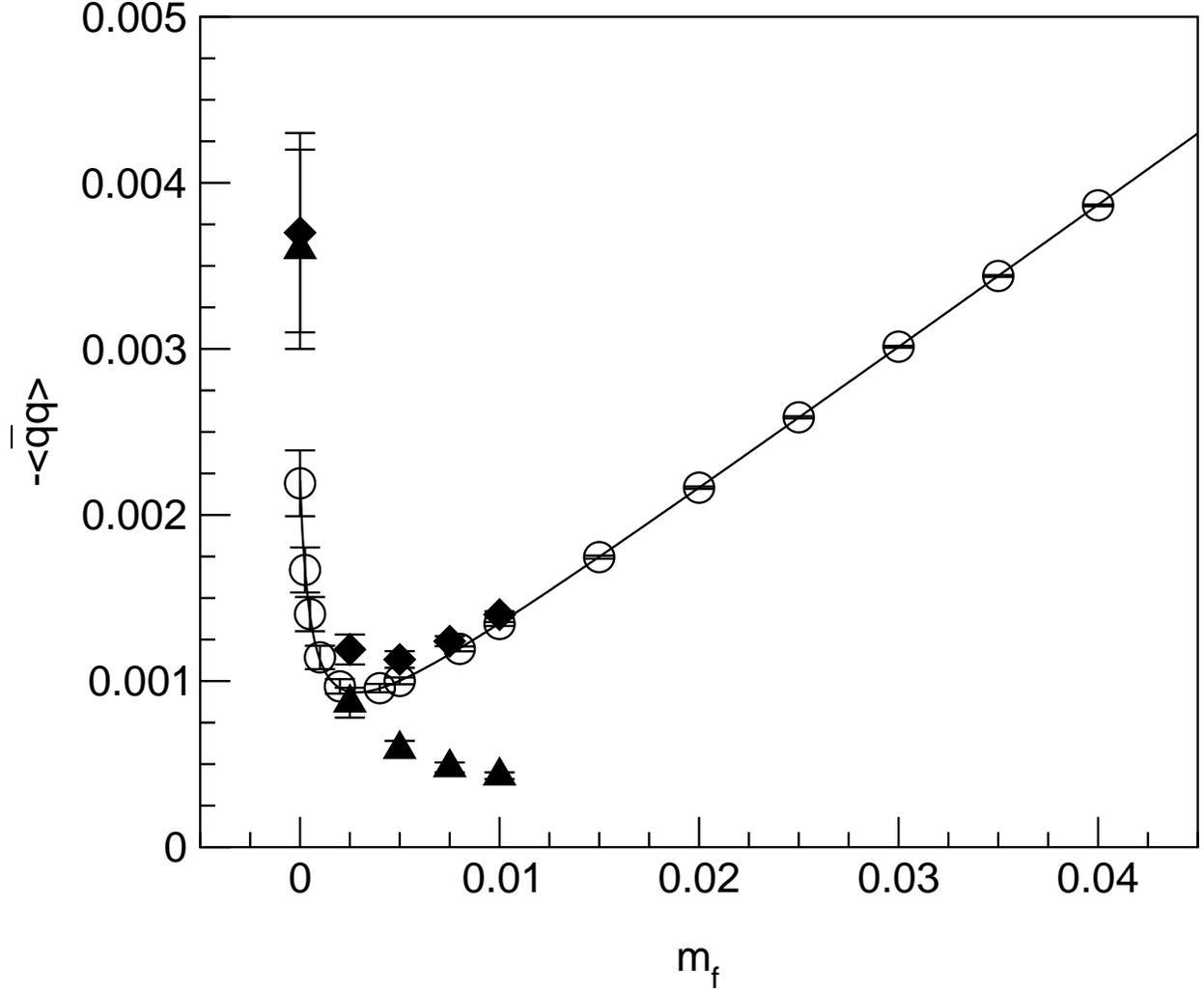}
\caption{$\qbq$ for quenched simulations done on $16^3 \times 32$
lattices at $\beta = 6.0$ for $L_s = 16$.  In addition to the direct
results for $\qbq$ presented in Ref.~\protect\cite{Blum:2000kn}, and
the fit of the form given in Eq.~\protect\ref{eq:qbq_fit}, we also 
plot the values obtained from the 18 eigenvectors studied here (filled
triangles) and the sum of those values with the contribution from
far off-shell states, $a_1(m_f+\mres)$ (filled diamonds).}
\label{fig:qbq}
\end{figure}

As a second consistency check, we investigate the extent to which the
lowest lying 18 eigenmodes which we examine actually play an important
role in the low energy physics described by the gauge configurations
being studied.  An easy way to address this question is to compute the
contribution of the modes which we have isolated to the chiral condensate
and to compare that contribution to the total chiral condensate determined
independently~\cite{Blum:2000kn}.  In Table~\ref{tab:qbq} and 
Fig.~\ref{fig:qbq} we present such a comparison.  As can be seen, these
lowest 18 eigenvectors provide a large fraction of the actual value of
$\qbq$ for the light input mass, $m_f=0$.  For larger values of $m_f$,
the contribution of these modes is a rapidly decreasing fraction of
$\qbq$.  However, for non-zero quark mass, the quadratically divergent
contribution of high mass states $\propto m_f/a^2$ should make an 
increasingly important contribution.  We can easily include these effects,
by using a fit to $\qbq$ of the form
\begin{equation}
  -\qbq = \frac{ a_{-1}}{ m_f + \dmqbq} + a_0 + a_1 m_f,
  \label{eq:qbq_fit}
\end{equation}
where $a_{-1}$, $a_0$, $a_1$ and $\dmqbq$ are parameters determined
and tabulated in Ref.~\cite{Blum:2000kn}.  The large coefficient
$a_1$ describes these divergent contributions, allowing them to be
included in our estimate of $\qbq$ by simply adding the term 
$a_1 \cdot (m_f+\mres)$ to $-\qbq$.  These results are included in 
Table~\ref{tab:qbq} and Fig.~\ref{fig:qbq}.  One sees that when this
expected contribution from far off-shell states is included, we
have good agreement with the directly computed values $\qbq$ for
$m_f$ in the range between 0.0 and 0.01.  This suggests that the 18
lowest modes that we have examined provide the bulk of the physical
vacuum expectation value $\qbq$ and are therefore relevant to an
attempt to characterize the physics of the QCD vacuum.

\begin{figure}
\epsfxsize=\hsize
\epsfbox{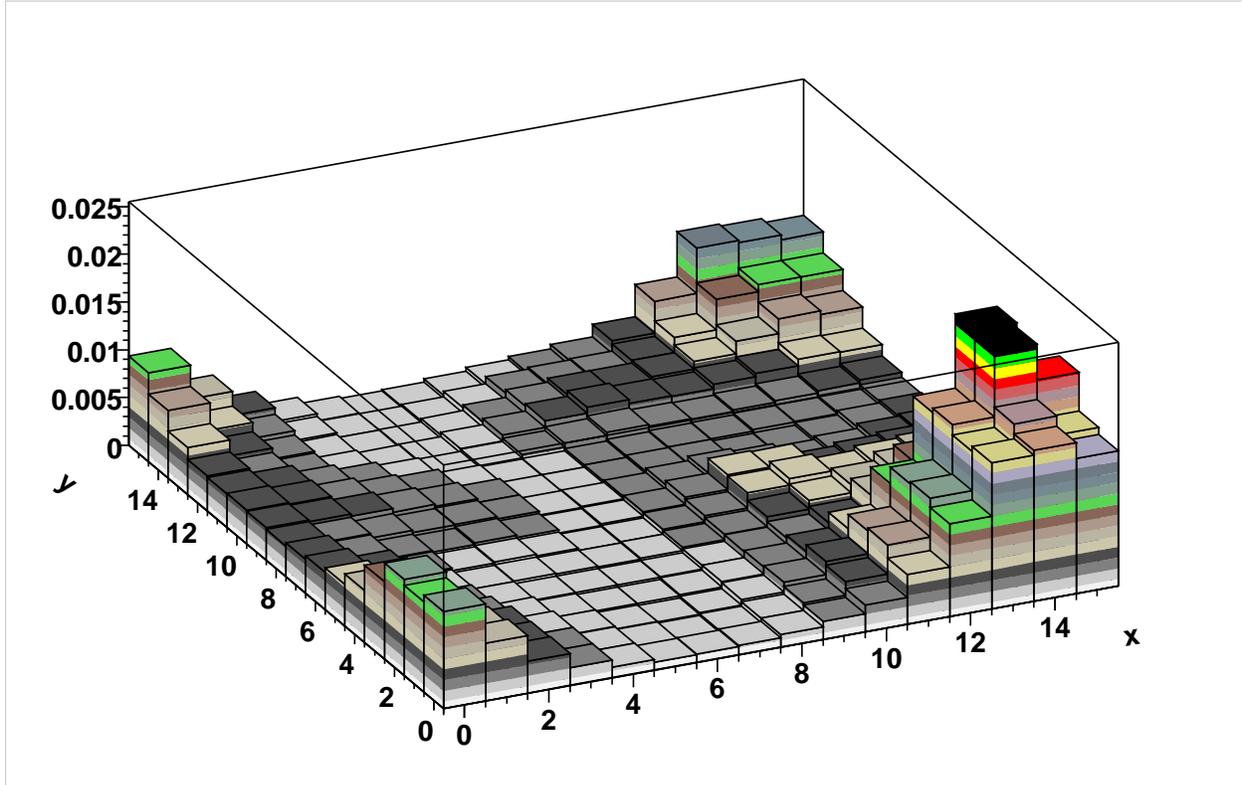}
\caption{The distribution of the magnitude of the right-handed components 
of a near zero mode eigenvector of $D_H$ in the $x-y$ plane, summed over
the $z$ and $t$ coordinates.  A single, well-localized peak is visible
near $x=14$ and $y=2-3$.}
\label{fig:evector_z_right}
\end{figure}

\begin{figure}
\epsfxsize=\hsize
\epsfbox{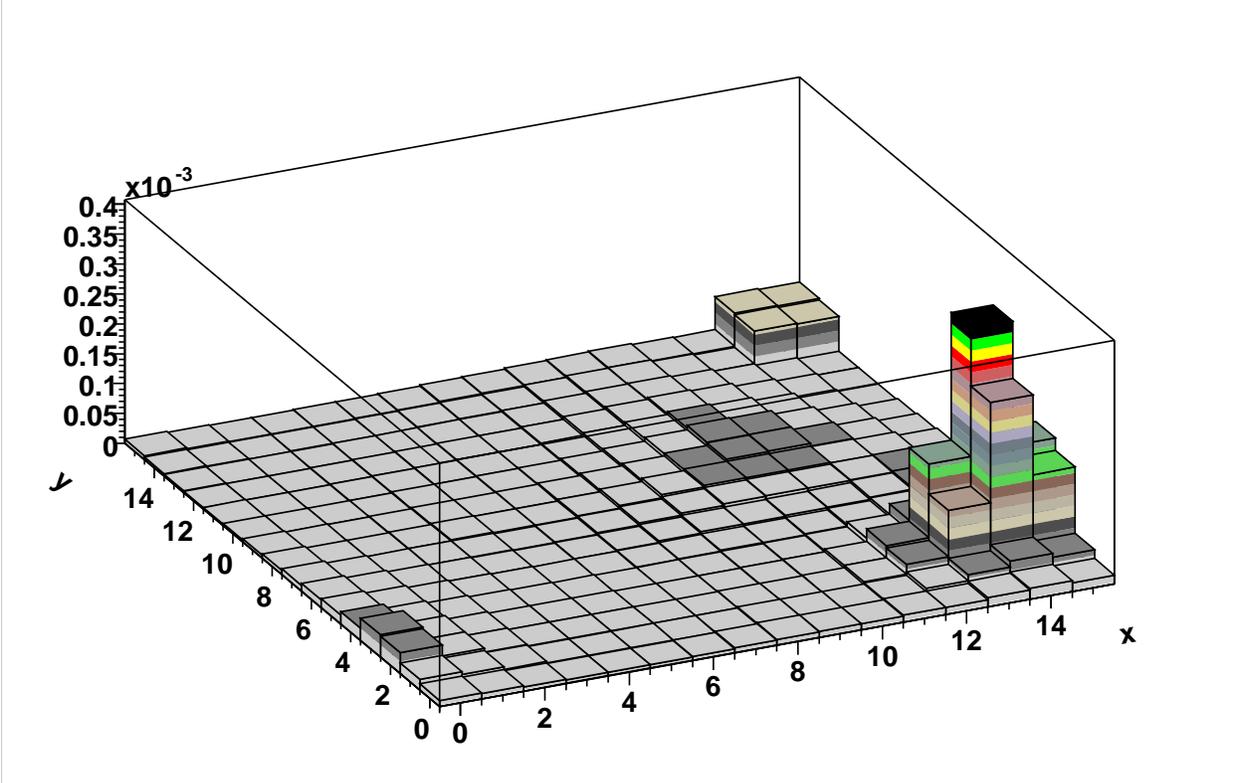}
\caption{The distribution of the magnitude of the left-handed components 
of a near zero mode eigenvector of $D_H$ in the $x-y$ plane, summed over
the $z$ and $t$ coordinates.  This is the same eigenvector whose right-handed
components are show in Fig.~\ref{fig:evector_z_right}.  While a peak can
also be seen in this figure, its magnitude is 50 times smaller than that 
shown for the right-handed components.}
\label{fig:evector_z_left}
\end{figure}

Finally, we note that the above results for the chiral density distribution 
of the near zero modes shown in Fig.~\ref{fig:local_chirality_zm} are not 
very illuminating since the local chirality is either left- or 
right-handed for all points in the lattice for these eigenvectors. 
A more informative picture is given in Fig.~\ref{fig:evector_z_right}
which shows the magnitude of the right-handed components of a near zero 
mode in the $x-y$ plane summed over $z$ and $t$. The mode is clearly 
localized (and remains so in the other planes that are not shown). 
It is also dominantly right-handed since the left-handed component shown
in Fig.~\ref{fig:evector_z_left} is more than 50 times smaller.  These 
figures corresponds to the unique, near zero mode labeled as 
eigenvector 0 in Fig.~\ref{fig:simple_lego}.  

\begin{figure}
\epsfxsize=\hsize
\epsfbox{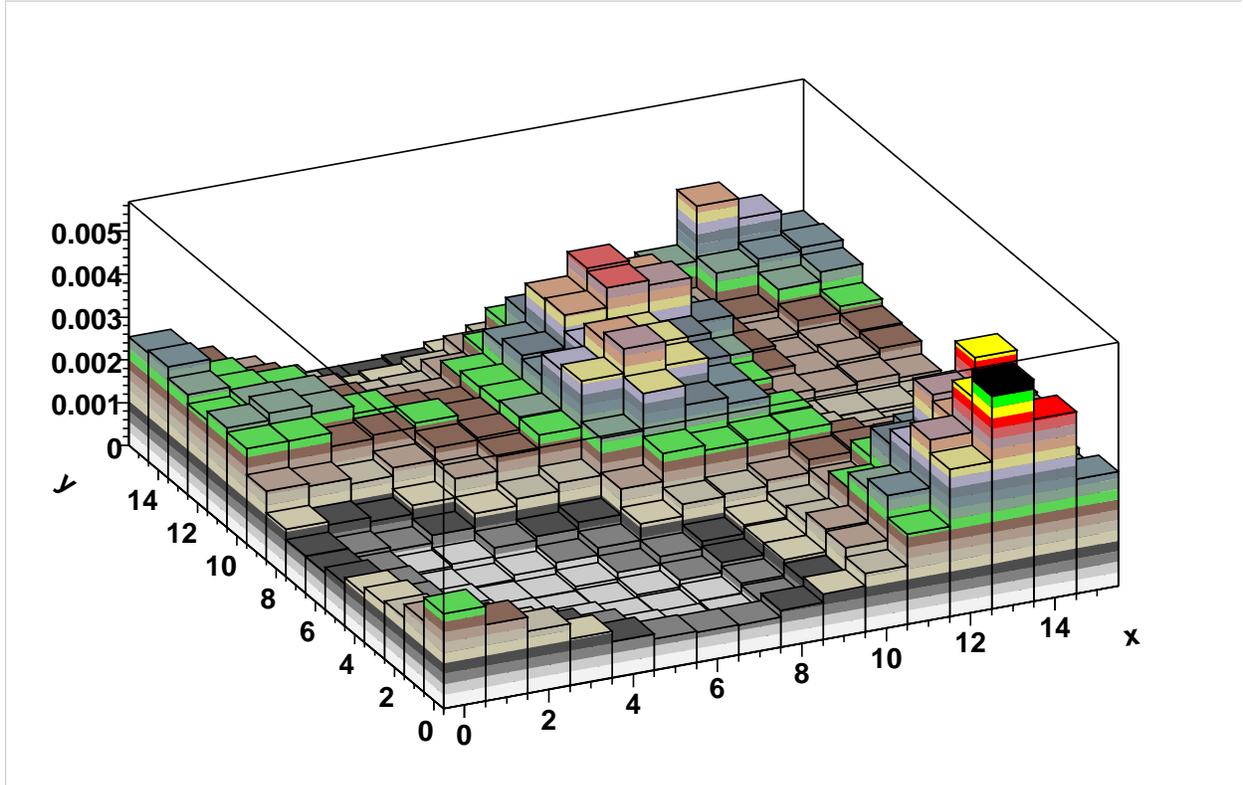}
\caption{The distribution of the magnitude of the right-handed components 
of a non-zero eigenvector of $D_H$ in the $x-y$ plane, summed over
the $z$ and $t$ coordinates.  A single somewhat broad peak is seen with
$x = 13$ and $y=0$ or 15.  [Note: it should be remembered that our periodic
boundary conditions imply that the points $y=0$ and $y=15$ are adjacent.]}
\label{fig:evector_nz_right}
\end{figure}

\begin{figure}
\epsfxsize=\hsize
\epsfbox{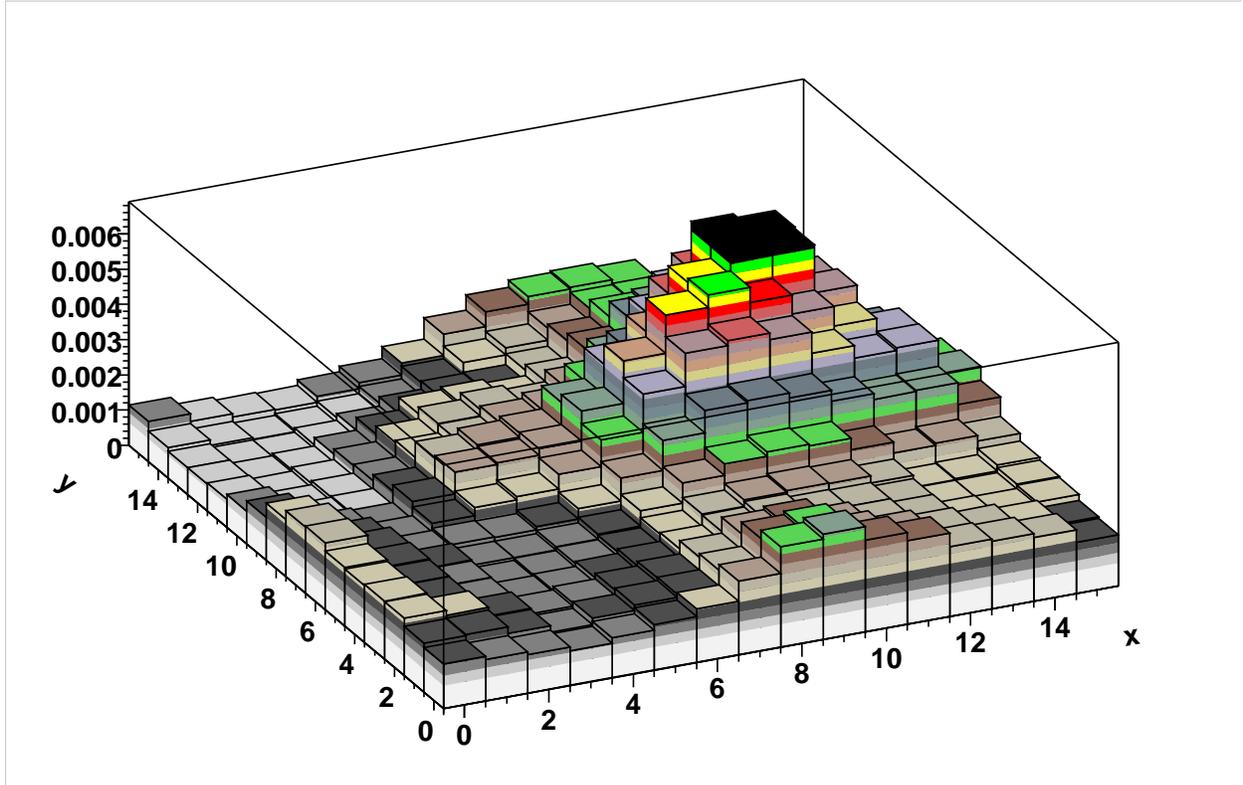}
\caption{The distribution of the left-handed components of the non-zero 
eigenvector shown in Fig.~\ref{fig:evector_nz_right}.  Note this shows
a quite different peak from that in the previous figure with $(x,y)=(11,11)$}
\label{fig:evector_nz_left}
\end{figure}

For completeness, we include similar figures showing the spatial 
distribution of the right- (Fig~\ref{fig:evector_nz_right}) and 
left-handed (Fig.~\ref{fig:evector_nz_left}) parts of the first 
non-zero mode, numbered 1 in Fig.~\ref{fig:simple_lego}.  These 
provide a particular illustration of the behavior shown in average 
terms in Fig.~\ref{fig:local_chirality_nzm}.  The right-handed 
components of eigenvector 1 are centered around the  point 
$(x,y,z,t)=(13,0,7,15)$ while the left-handed components of this
eigenvector are largest at the distinct point $(11,9,3,6)$.  This 
specific case shows a clear correlation between localization and 
chirality as would be expected in an instanton picture.

\section{Conclusion}
\label{sec:conclusion}

We have examined the local correlation between the chirality and magnitude
of the low-lying eigenmodes of the Dirac operator in quenched lattice QCD.
A strong correlation is observed in our calculation suggesting a picture
of the QCD vacuum containing background gauge fluctuations producing
space-time regions in which the Dirac eigenstates are both localized 
and chiral.  For those 0.7\% of the lattice sites at which 6\% of the 
eigenmode magnitude is concentrated, these eigenmodes are nearly 
eigenstates of $\gamma^5$ with $X_H(x)/\Omega_H(x)$ within 90\%
of 1.  While this is certainly consistent with a vacuum described as 
a superposition of instantons, it is an important feature that may be 
accommodated by other models as well.

An essential ingredient of our calculation is the use of the domain
wall fermion formulation.   The domain wall Dirac operator readily 
identifies topological zero modes on gauge configurations as 
they appear in the Feynman path integral without resort to smoothing 
or cooling.  The presence of these near zero modes in the domain
wall Dirac spectrum is easily seen in the small mass behavior of the 
chiral condensate~\cite{Fleming:1998cc} and their implications for 
quenched calculations of the both the chiral condensate and hadron 
masses was discussed in detail in Ref.~\cite{Blum:2000kn}.
Related behavior has been found using the overlap Dirac 
operator~\cite{Dong:2000mr,DeGrand:2000gq}, an alternative lattice 
fermion formulation, also with improved chiral symmetry.

The recent calculation~\cite{Horvath:2001ir} using Wilson fermions 
which motivated this study found no chirality-localization 
correlation.  However, the calculation of the present paper is 
improved over this earlier work in two significant respects.  The 
most important is that we use domain wall fermions which preserve 
chiral symmetry to a high degree at non-zero lattice 
spacing~\cite{Blum:2000kn,AliKhan:2000iv}, while Wilson fermions do 
not.  In addition the lattice spacing used here is roughly twice as 
small as that of Ref.~\cite{Horvath:2001ir}.

Note added: After this paper was essentially complete, the recent
work of DeGrand and Hasenfratz became available~\cite{DeGrand:2001pj}.
Our analysis and conclusions are quite similar to those of this paper.
While both that paper and the present one, use improved fermion
Dirac operators, overlap and domain wall respectively, we have
used the gauge configurations directly, without smearing or fattening.
This suggests that this strong correlation between spacial localization 
and chirality is seen even for gauge configurations whose fluctuations 
span the full range of distance scales produced by both the Iwasaki 
and Wilson gauge actions.

\section*{Acknowledgments}

We thank RIKEN, Brookhaven National Laboratory and the U.S. Department 
of Energy for providing the facilities essential for the completion 
of this work.  We also thank Robert Edwards for providing us with his
Ritz diagonalization program.

The numerical calculations were done on the 400 Gflops QCDSP 
computer~\cite{Chen:1998cg} at Columbia University and the 600 Gflops 
QCDSP computer~\cite{Mawhinney:2000fx} at the RIKEN BNL Research Center.  
This research was supported in part by the DOE under grant 
DE-FG02-92ER40699 (Columbia), in part by the DOE under grant
DE-AC02-98CH10886 (Dawson), and in part by the RIKEN-BNL Research
Center (Blum).

\bibliography{eigen}

\begin{table}
\caption{\label{tab:cuts} 
Fractions of lattice sites and eigenvector normalization included for 
the six cuts, $\Omega_H(x) > \Omega_{\rm min}$, used in this paper.}
\begin{tabular}{ccccc}
                   & \multicolumn{2}{c}{Iwasaki Action} &
                              \multicolumn{2}{c}{Wilson Action} \\
\hline
$\Omega_{\rm min}\cdot10^5$ 
                   & Fraction of   & Fraction of & Fraction of   & Fraction of   \\
                   & lattice sites & normalization 
                                                 & lattice sites & normalization\\ 
\hline
3   & 0.086   & 0.285   & 0.080  & 0.319 \\
4   & 0.043   & 0.186   & 0.042  & 0.235 \\	
5   & 0.024   & 0.132   & 0.026  & 0.186 \\
6   & 0.015   & 0.100   & 0.017  & 0.156 \\
7   & 0.010   & 0.079   & 0.012  & 0.136 \\
8   & 0.007	  & 0.064   & 0.009  & 0.121 \\
\end{tabular}
\end{table}
\begin{table}
\caption{\label{tab:qbq} 
Contributions to the chiral condensate from the 18 lowest eigenvectors 
examined in this paper, $\qbq_{18}$, for five different quark mass values. 
In addition, we include the $\sim 1/a^2$ contribution from the large
eigenmodes by including the term $a_1(m_f+\mres)$ using the 
values $a_1= 3.97(8) 10^{-4}$ and $\mres=1.24(5) 10^{-3}$ 
taken from Ref.~\protect\cite{Blum:2000kn}.  Finally these results are 
compared with the complete value for $\qbq$ found in that paper.  This 
comparison is done in the case of the Wilson action at $\beta=6.0$ 
where the detailed results for $\qbq$ are available from that 
earlier work.}
\begin{tabular}{cccc}
$m_f$   & $-\qbq_{18}$  & $-\qbq_{18}+a_1(m_f+\mres)$ & $-\qbq$ \\
\hline 
0.0     & 0.0036(6)    & 0.0037(6)                  & 0.00219(20)\\
0.0025  & 0.00087(9)   & 0.00119(9)                 & -          \\
0.005   & 0.00059(5)   & 0.00113(5)                 & 0.00100(2) \\
0.0075  & 0.00048(3)   & 0.00124(3)                 & -          \\
0.01    & 0.00043(2)   & 0.00140(2)                 & 0.00134(1)
\end{tabular}
\end{table}

\clearpage

\clearpage

\clearpage

\clearpage

\clearpage


\clearpage

\clearpage

\clearpage

\clearpage

\clearpage

\end{document}